\documentclass{aa}
\usepackage{graphicx}
\begin{document}
\title{Astrometry with {\it Carte du Ciel} plates, San Fernando zone. \\
II. CdC-SF: a precise proper motion catalogue\footnote{The catalogue 
is only available in electronic form at the CDS 
via anonymous ftp to cdsarc.u-strasbg.fr (130.79.128.5) or 
via http://cdsweb.u-strasbg.fr/cgi-bin/qcat?J/A+A/}
}

\author{B. Vicente \inst{1,2},
C. Abad \inst{3},
F. Garz\'on \inst{1,4}
\and
T.~M. Girard \inst{5}}

\offprints{B. Vicente (bvicente@iaa.es)}

\institute{Instituto de Astrof\'{\i}sica de Canarias (IAC),
La Laguna (S/C de Tenerife), E-38200 Spain.
\and
Current address: Instituto de Astrof\'{\i}sica de Andaluc\'{\i}a
(IAA-CSIC), Apdo. 3004, E-18080, Granada, Spain.
\and
Centro de Investigaciones de Astronom\'{\i}a (CIDA),
Apdo. 264, M\'erida, 3101-A Venezuela.
\and
Departamento de Astrof\'{\i}sica, Universidad de La Laguna,
La Laguna (S/C de Tenerife), E-38200 Spain.
\and
Department of Astronomy, Yale University, P.O. Box 208101, 
New Haven, CT 06520, USA
}

\date{Received 9 July 2009 / Accepted 5 November 2009}

% \abstract{}{}{}{}{}
% 5 {} token are mandatory

\abstract
% context heading (optional), leave it empty if necessary
{The historic plates of the {\it Carte du Ciel}, an international cooperative
project launched in 1887, offer valuable first-epoch material for the
determination of absolute proper motions.
}
{We present the CdC-SF, 
an astrometric catalogue of positions and proper motions
derived from the  \textit{Carte du Ciel} plates of the San Fernando zone,
photographic material with a mean epoch of 1901.4 and a limiting magnitude of
V$\sim$16,
covering the declination range of $-10^{\circ} \leq \delta \leq -2^{\circ}$.
}
{Digitization has been made using a conventional flatbed scanner.
Special techniques have been developed to handle the combination of plate
material and the large distortion introduced by the scanner.
The equatorial coordinates are on the ICRS defined by Tycho-2, and
proper motions are derived using UCAC2 as second-epoch positions.
}
{
The result is a catalogue with positions and proper motions for 560000 stars,
covering 1080 degrees$^2$. The mean positional uncertainty is 0$\farcs$20
(0$\farcs$12 for well-measured stars) and the proper-motion
uncertainty is 2.0 mas/yr
(1.2 mas/yr for well-measured stars).
}
{The proper motion catalogue CdC-SF is effectively a deeper extension
of Hipparcos, in terms of proper motions, to a magnitude of 15.
}


\authorrunning{Vicente, B. et al.}
\titlerunning{CdC-SF Catalogue}

\maketitle

%
%________________________________________________________________

\section{Introduction}

The Hipparcos Catalogue (Perryman et al. 1997) contains the positions, 
parallaxes and proper motions of 118218 stars with great precision
(proper motions precision is 1~mas/year), complete up to a magnitude of $V=7.3$.
Due to its precision and homogeneity, 
this catalogue constitutes an important source of information and 
has provided a useful tool in numerous kinematic studies.
However, its relatively bright magnitude limit
restricts it to very nearby stars, with the number of stars
per unit area also being lower than what is often desired in kinematic
analyses of the Galaxy.

For this reason, various efforts have been made to develop 
astrometric catalogues that are more dense and reach deeper magnitudes, 
becoming in effect an extension of the Hipparcos catalogue.
It is necessary that such catalogues contain proper motions 
besides positions in order to provide the essential 
information with the quality and completeness needed.

The Tycho-2 Catalogue (H{\o}g et al. ~ 2000) can be considered
an extension of Hipparcos in the way that it contains positions and
proper motions for stars brighter than V$\sim$12 with errors of 2.5~mas/year. 
At the present time the Hipparcos and Tycho-2 catalogues are the largest
high-quality proper motion surveys.

Other proper motion catalogues available at the present time
are UCAC2 catalogues (Zacharias et al. 2004, 2--7 mas/yr depending
on the magnitude and current version has systematic errors 
of the order of 3~mas/yr),
NPM (Hanson et al. 2004, 5~mas/yr, $8<B<18$) 
and SPM (Girard et al. 2004, 4~mas/yr, $5<V<18.5$).
For an overview in detail of the current astrometric catalogues, see Girard (2008).

When determining proper motions, a long time-baseline is highly desirable.
Historical photographic records provide ideal material for the calculation of 
proper motions. 
This is a prime motivation for the digitization and 
reduction of the collections of photographic plates that were accumulated 
in observatories at the end of the 19th century.

In 1887, the first great astronomical project of international cooperation 
in history was planned in Paris. Its objective was
the construction of a complete catalogue up to a magnitude of V$\sim$12,
the {\it Astrographic Catalogue} (AC), and to map the sky up to V$\sim$15,
the {\it Carte du Ciel} (CdC). 
A total of twenty observatories around the world participated in
the task of taking the necessary observations.
This material represents the first existing photographic record of the 
complete sky and is an extremely important potential resource for
determining proper motions due to the long time-baseline relative to
modern observations,
$\sim$100 years in most cases.

This study is part of a larger project, the goal of which is to produce 
an astrometric catalogue with proper motions for a variety of applications
in Galactic kinematics, as well as to provide a
dense, faint reference frame by recovering and utilizing the historical CdC plates.
In order to construct the catalogue, the first necessary step is 
to digitize these plates.
The digitization and measurement process is discussed in detail in a previous paper 
(Vicente at al. 2007, hereafter Paper~I), where 
an innovative method of digitization using a commercial flatbed scanner was presented. 
It was demonstrated that this device, together with the techniques developed 
to remove the distortion it introduces, is an alternative 
to more specialized measuring machines that are less available, while 
yielding similar measuring errors of 0$\farcs$18 as
other groups, deriving astrometry from similar plate material, 
although from other CdC collections (Rapaport 2006, Ortiz-Gil 1998, Geffert 1996).

The main scientific value of positions at the epoch of 1900
resides in their usefulness in providing positions to serve as 
the first epoch in proper motion determinations involving the compilation of 
several catalogues of different epochs.

The current paper explains the various steps and procedures used
during the construction of the CdC-SF proper-motion catalogue.
The catalogue's properties are described in detail, including an 
estimation of uncertainties in the positions and proper motions.
For the convenience of the reader, 
some of the figures and tables from Paper~I are reproduced here
for their relevance to the catalogue description.

\section{Digitization of the plate material with a flatbed scanner}

The plate material used in this work corresponds to the 
CdC collection stored in the Real Instituto y Observatorio
de la Armada in San Fernando, Spain (ROA), which was charged with
observing the area between the -2$^{\circ}$ and -10$^{\circ}$ declination.
This collection of 1260 {\it Carte du Ciel} plates has not been
exploited up to now.
Each plate covers a field of $2^{\circ} \times 2^{\circ}$, and
observations were planned in a full overlapping strategy.
Plates along odd declinations were exposed three times,
producing a pattern of images for each star that is
roughly an equilateral triangle.
All of the plates also contain superposed {\it r\'eseau} grid lines.

The plates are a historical patrimony of the institution, 
and as such they could not be extracted from the museum to
be measured with specialized instrumentation.
Initially, reproductions were made in acetate and these were to be digitized
with a measuring machine, a microdensitometer PDS.
After performing several test scans of the acetate copies using 
the PDS at Yale University (USA) 
and the PDS at the CIDA (Venezuela), it was concluded
that large unpredictable distortions due to the copying process
(up to 15~$\mu m$) 
together with the slowness of the PDS (1 plate / day) 
would prohibit the successful completion of the project.

For this reason, we used a commercial flatbed scanner as an alternative for 
the digitalization.
Though its internal precision is much inferior to that obtained 
from a PDS,  
it is possible to measure the original plates with the scanner
and, because of its faster speed (1 plate/8 min.), 
it allowed us to repeat the digitalization to improve the final precision.
At present, ROA has completed the digitization of its
collection of 2520 AC/CdC plates.

\begin{figure}
\centering
\includegraphics[width=0.45\textwidth]{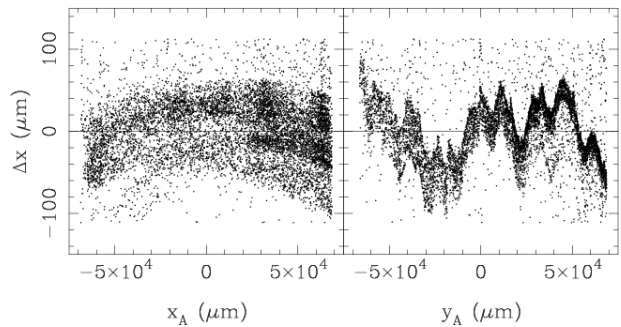}
\caption{
Positional differences obtained from two digitalizations of the same plate, 
rotated one with regard to the other by 90$^{\circ}$ and then 
aligned via a linear transformation.
Notice that the amplitude in the errors in the $Y$ axis is considerable. 
In addition, this distortion also varies from scan to scan and from plate to plate.
}
\label{fig01}
\end{figure}

The main problem to be addressed when using a measuring device such as a 
scanner for astrometric work is the significant distortion introduced 
($\sim$100~$\mu m$, Fig.~\ref{fig01}) by the scanner itself.
To make matters worse, this distortion is not consistent, as
has been confirmed by repeated sans.
Vicente et al. 2007 described the method developed to calibrate and
correct for scanner distortion in detail yielding excellent results.
The resultant final single-measurement internal error-estimate per exposure is
3~$\mu$m (0$\farcs$2) for single-exposures plates
and 5~$\mu$m (0$\farcs$3) for triple-exposures ones.

\section{Measurement of the plates}

The most important characteristics
of the plates that complicate the determination of precise
astrometry include:
1) the blending of the triple-exposure images on the odd-numbered
declination plates,
2) the blending and confusion of stars that fall on the superposed 
{\it r\'eseau} grid lines,
3) the false detections due to plate flaws, spurious dust
and degradations that have accumulated during storage,
and
4) typical effects in photographic material caused by optical aberrations
and by non-linear detectors.

A loss of up to 15\% of stars
can be expected due to interferences with the grid lines and
spurious flaws as well as to the blending of the triple exposures.

For the resolution of all these problems it has been necessary to develop 
the necessary programmes from scratch
since available commercial software packages 
are not optimized for photographic material.
All of the effects mentioned above have been successfully treated.  Details
can be found in Paper~I.

Currently, the algorithms have been applied to only
one third of the San Fernando zone; 
the one with a right ascension between 6 and 
14 hours, constituting 420 plates (180 simple-exposures plates and 240
triple-exposures plates).
The measurement of the remaining plates in the future will be
straightforward, based on the information developed and presented here
and in Paper~I.

\section{Astrometric reduction of the plates }

  The transformation from the $(x,y)$ coordinates into celestial
coordinates ($\alpha,\delta$) was performed by the
block-adjustment technique (Stock~1981)
including a determination of the field distortion (Abad~1993).
This technique utilizes images which overlapping plates have in common
in addition
to positional information from an external reference catalogue.
A linear plate model is used in combination with a residuals-based
corrective mask that is common to all plates.  
These are derived and applied in an iterative process.

\begin{figure}[h!]
\centering
\includegraphics[width=0.45\textwidth]{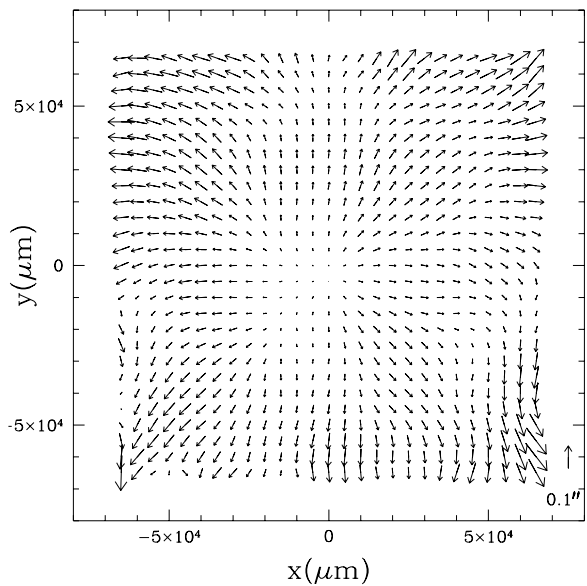}
\caption{Position residuals as a function of coordinates after a linear plate
model is applied, showing the systematic errors in the plates
due to the telescope optics.}
\label{fig02}
\end{figure}

Residuals consist of the differences between individual positions and the 
averages, for multiply measured stars, in addition to differences between 
the averages and the external catalogue for the reference stars.
Figure~\ref{fig02} shows a sample of the stacked residuals.  The pattern 
represents the systematic field distortion remaining in the plates. 
This empirical function is applied to the positions, and a new
iteration of the linear plate solution plus mask is performed.
For more details about the procedure and other considerations in the
determination of positions see Paper~I.
The overlapping technique allows for the simultaneous reduction of all plates.
Nevertheless, we decide to divide the set of plates 
into four zones grouped by right ascension in order to compare
the results as a function of the sky position and stellar density.
Every zone covers two hours in right ascension
and seven complete zones of declination.

The Tycho-2 Catalogue (H{\o}g et al.~2000) was used as a reference catalogue,
using its proper motions to take the positions backward to the epoch 
of each plate.  Its proper motions have a precision of 2.5~mas/year and
an accuracy somewhat less than that, making it a suitable reference source. 
It also has a sufficient density of stars and a magnitude limit of V$\sim$11.5.
An alternative reference catalogue, the UCAC2, was not adopted due to its 
lower precision in proper motions,
and also because its positions will be used as second-epoch data
in the proper motion determination, and we prefer a totally independent
reduction of the first-epoch material.

\subsection{Magnitude-equation correction}

\begin{figure}[ht!]
 \centering
 \includegraphics[width=0.4\textwidth]{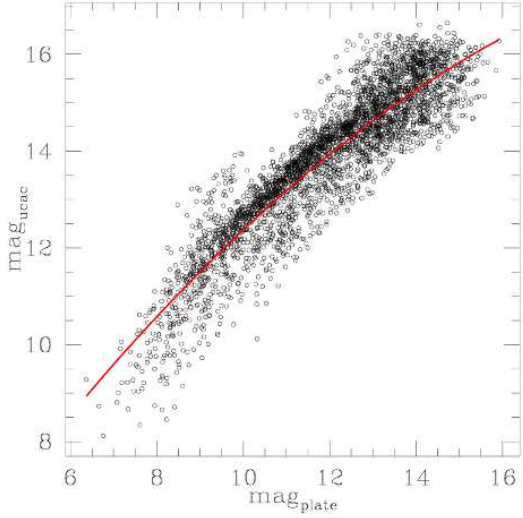}
 \caption{Example of the magnitude calibration for one plate.
A second-order polynomial fitting is applied (red line).}
 \label{fig03}
 \end{figure}

Magnitude-equation is the term used to describe the bias in star image
positions as a function of stellar magnitude that is present in virtually
all photographic plates.
In order to correct for it, it is necessary to obtain not only
the Cartesian positions 
$ (x, y) $, of each stellar image but also a magnitude index for the image,
to allow a magnitude calibration.
The estimated magnitude we use is the photographic one, 
the measured flux under the 2D Gaussian fitting,
calibrated with the approximate $R$ magnitudes of the UCAC2 catalogue.
For each plate we calculate the coefficients of the transformation needed 
as a least-squared fitting with a second-order polynomial (Fig.~\ref{fig03}).

\begin{figure}[ht!]
\centering
\includegraphics[width=0.45\textwidth]{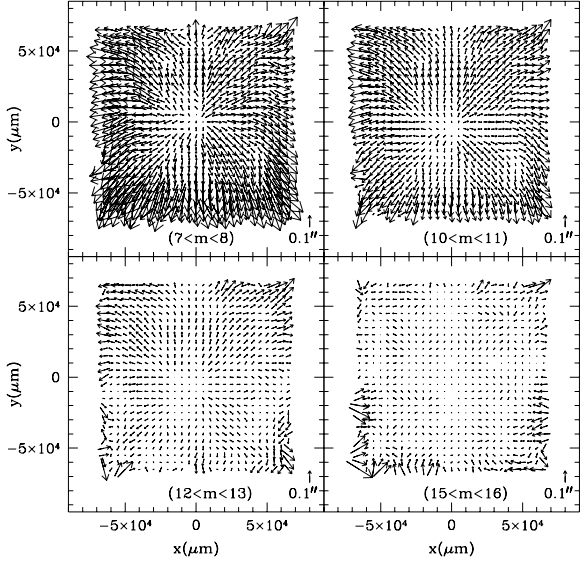}
\caption{Vector residuals are shown for four exemplary magnitude
ranges. 
Positions within a given magnitude range are corrected using the 
appropriate mask. 
The exact correction for a star is calculated as an interpolation in 
magnitude between the two masks nearest in magnitude.} 
\label{fig04}
\end{figure}

The common magnitude equation within the plates is handled by stacking
residuals and deriving correction masks. 
Different correction masks are constructed by binning stars into 
one-magnitude wide intervals to determine the magnitude dependence 
of the systematic errors (some examples in Fig.~\ref{fig04}).
The distortion is found to be more pronounced at bright magnitudes.

\begin{figure}[h!]
\centering
\includegraphics[width=0.45\textwidth]{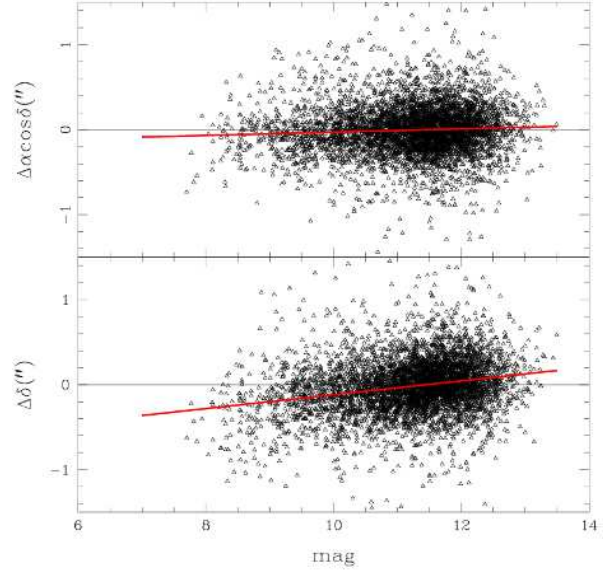}
\caption{
Systematic trend in the positional differences for Tycho-2 reference stars 
depending on the magnitude.
The straight line is a least-squares linear fit.
These data correspond to the group of the plates with 
$06^h \leq \alpha \leq 08^h$, as an example.}
\label{fig05}
\end{figure}

After the mask correction is applied, 
a slight residual magnitude equation was
detected based on differences with the
reference Tycho-2 catalogue (Fig.~\ref{fig05}).
However, it is possible that these systematic effects are actually
inherent in the Tycho-2 proper motions and propagated into the Tycho-2
positions at epoch 1900. 
After a thorough study of our derived  
proper motions for open cluster members, we decided for this to be the case 
and did not make a further adjustment in the magnitude-equation based on
the trends seen in the differences with Tycho-2.

\section{Determination of the proper motions}

The early-epoch CdC positions are combined with modern
positions from the UCAC2 catalogue to derive relative proper motions.
We have confirmed previously that both catalogues 
are free of significant systematic errors with respect to Tycho-2, 
minus the residual magnitude trend discussed above.
Thus, both are ostensibly on the same system.

The relative proper motions are placed on the ICRS system via
a direct comparison to the Hipparcos proper motions for common stars
with our catalogue, applying a local correction for systematic effects.
The value to be corrected is calculated as an average
of the differences between our proper motions and those of the 50 nearest 
Hipparcos stars.
We have preferred using the accumulative distribution function 
(Hamaker 1978) to calculate the average and the dispersion of the 
Hipparcos proper-motion differences for the 50 nearby stars.
In general, this method is much more 
consistent and immune to atypical values for the estimation of the average 
and dispersion using the property of which the sample fits 
a Gaussian distribution.

\begin{figure}[h!]
\centering
\includegraphics[width=0.45\textwidth]{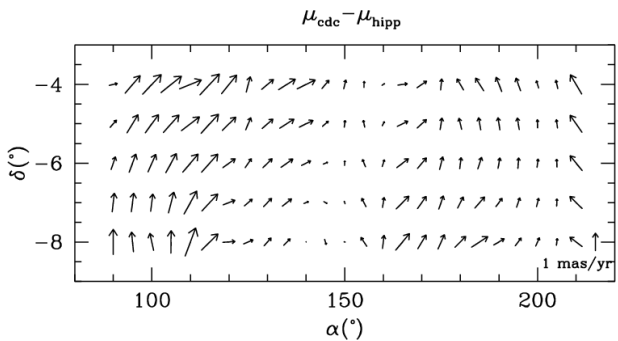}
\caption{Differences in the proper motions of CdC-SF and Hipparcos
as a function of right ascension and declination before
the local correction has been applied.
After the application the differences become random and less than 1 mas/yr.}
\label{fig06}
\end{figure}

Figure~\ref{fig06} shows a vector field with 
the differences in proper motions between the two sets of proper motions
as a function of sky position, employing for the plot 
the technique of the ``weighted sliding polynomial'' (Stock \& Abad 1988).
The source of the systematic trend seen in the differences between the 
preliminary CdC-SF proper motions and those of Hipparcos is unknown.  
It is difficult to explain trends with a scale length of tens of degrees 
when our individual plates cover just two degrees.  
We speculate that it may be due to systematics in the Tycho-2 proper motions 
that were used in deriving our first-epoch positions.  
Systematics of this amplitude, of 1 to 2 mas/yr, have been noticed during the construction 
of the soon to be released Southern Proper Motion catalogue (SPM4) when compared to Tycho-2.
Noticeable fixed-pattern trends of 1 to 2 mas/yr in amplitude and a scale length of one to two degrees 
were observed across the six-degree wide SPM fields, even though Tycho-2 had been used as a reference 
in the cubic polynomial plate models.  Whether larger scale-length systematics are present 
in the Tycho-2 proper motions is not known (and could not have been detected during the 
SPM4 analysis, as they would have been absorbed by the cubic polynomial), 
but this is a possibility (Girard 2009, private comm.).
It is also demonstrated in Rapaport (2006) that the propagated Tycho-2 errors are non-negligible, 
becoming equal to or even greater than the measurement errors
due to the long interval of time between the epochs of Tycho-2 positions and 
CdC plates (about 100 years).
For now, the cause of the pattern seen in Fig.~\ref{fig06} is unclear.  
Nevertheless, the best course of action is to correct for it as we have done, 
placing us more firmly on the proper-motion system of Hipparcos, 
which we consider to be reliable.

We have also confirmed that the magnitude equation is not present
for the common stars with Hipparcos.
Thus we conclude that our absolute proper motions are indeed in the Hipparcos
system, at least for the common magnitude range.

\section{Error estimation}

As an example of the astrometric quality of the final catalogue
which we have extensively described,
we will present here the errors and uncertainties of its positions
and proper motions.

\subsection{Errors in positions}

An estimated internal error for each star in the catalogue is derived based on
the rms of the positional differences of each image in the overlapping plates
to the average position.
This value, considered as the internal positional error of the star, 
is included in the catalogue.
The distribution of the errors 
as a function of magnitude is given in Table~\ref{table01},
repeated here from Paper~I for the convenience of the reader.

\begin{table*}[ht!]
 \caption{Final internal positional uncertainties as a function of
magnitude for the derived CdC-SF coordinates, grouped by right ascension.
Also listed are the standard deviations of differences with the Tycho-2 
positions at the epoch of the plates.}
 \label{table01}
 \centering
% \[
 \begin{tabular}{|r|rrr|rrr|rrr|rrr|}

 \noalign{\smallskip}
 \hline
 \noalign{\smallskip}

 \multicolumn{1}{|c|}{Area} & \multicolumn{3}{c|}{$06^h\leq\alpha<08^h$} & \multicolumn{3}{c|}{$08^h\leq\alpha<10^h$} &
 \multicolumn{3}{c|}{$10^h\leq\alpha<12^h$} & \multicolumn{3}{c|}{$12^h\leq\alpha<14^h$} \\

 \noalign{\smallskip}
 \hline
 \noalign{\smallskip}

 Magnitude & $\sigma_{\alpha} ('')$ & $\sigma_\delta ('')$ & $N_{stars}$ & 
       $\sigma_{\alpha} ('')$ & $\sigma_\delta ('')$ & $N_{stars}$ &
       $\sigma_{\alpha} ('')$ & $\sigma_\delta ('')$ & $N_{stars}$ &
       $\sigma_{\alpha} ('')$ & $\sigma_\delta ('')$ & $N_{stars}$ \\

 \noalign{\smallskip}
 \hline
 \noalign{\smallskip}

    8 &  0.10 &  0.11 &      146 &  0.12 &  0.10 &      114 &  0.12 &  0.16 &       43 &  0.08 &  0.09 &       65 \\
    9 &  0.11 &  0.11 &     1066 &  0.11 &  0.10 &      710 &  0.11 &  0.12 &      369 &  0.11 &  0.10 &      361 \\
   10 &  0.12 &  0.12 &     5437 &  0.12 &  0.11 &     3009 &  0.13 &  0.12 &     1616 &  0.13 &  0.11 &     1368 \\
   11 &  0.11 &  0.10 &    16786 &  0.11 &  0.10 &     7414 &  0.13 &  0.12 &     3445 &  0.13 &  0.11 &     3193 \\
   12 &  0.10 &  0.10 &    34489 &  0.11 &  0.10 &    13918 &  0.13 &  0.12 &     5691 &  0.13 &  0.13 &     5077 \\
   13 &  0.11 &  0.11 &    61426 &  0.12 &  0.11 &    24092 &  0.15 &  0.14 &     8714 &  0.15 &  0.14 &     7843 \\
   14 &  0.22 &  0.20 &   119957 &  0.23 &  0.21 &    46563 &  0.24 &  0.22 &    15161 &  0.26 &  0.24 &    15267 \\
   15 &  0.32 &  0.31 &    84942 &  0.32 &  0.30 &    30315 &  0.31 &  0.30 &     8916 &  0.33 &  0.31 &     9042 \\

 \noalign{\smallskip}
 \hline
 \noalign{\smallskip}

Total &  0.21 &  0.19 &   324335 &  0.20 &  0.19 &   126142 &  0.21 &  0.19 &    43960 &  0.22 &  0.21 &    42227 \\

 \noalign{\smallskip}
 \hline
 \noalign{\smallskip}

$\Delta$ Tycho-2  &  0.30 &  0.33 & 19066 &  0.34 &  0.35 & 9127 &  0.42 &  0.34 & 4168 &  0.37 &  0.33 & 4028 \\

 \noalign{\smallskip}
 \hline
 \noalign{\smallskip}
 \end{tabular}
% \]
\end{table*}

The mean values of these uncertainties are
$(\sigma_{\alpha cos \delta},\sigma_{\delta})=(0\farcs21,0\farcs19)$
for the entire catalogue, and for stars brighter than 14, the mean values are
$(\sigma_{\alpha cos \delta},\sigma_{\delta})=(0\farcs12,0\farcs11)$.
For completeness, triple exposure plates with a measuring error of better
than 7$\mu m$ are included in the catalogue. 
Nevertheless, for minimizing the effect of their lower precision, 
a lesser weight has been assigned to them.
The mean error of the catalogue including 
only simple exposure plates is 0$\farcs$14, in each coordinate.
The individual uncertainty estimates are based on a handful of 
measures per star. Therefore, one should not place too much weight 
on the variation of uncertainty estimates within a given magnitude range, 
and the median uncertainty at a given magnitude may possibly be a better estimate for any
particular star.

The final line in Table~\ref{table01} also lists the rms differences 
between our derived positions and the Tycho-2 catalogue, at the CdC-SF epoch.
These rms differences are expected to have a significant, possibly dominant,
contribution from the errors in the Tycho-2 positions at this epoch.

\begin{figure}[h!]
 \centering
 \includegraphics[width=0.4\textwidth]{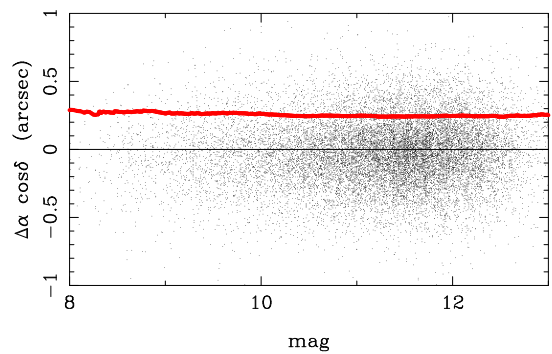}
 \caption{Differences in positions between CdC-SF positions and Tycho-2
 catalogue at the CdC epoch as a function of magnitude for stars in
 common. The fitted line indicates the moving-mean standard
 deviation. The plot for $\Delta\delta$ is similar.} 
 \label{fig07}
 \end{figure}

A comparison of our catalogue with the Tycho-2 positions at the CdC plates' epoch
as a function of magnitude (Fig.~\ref{fig07}) 
shows no systematic pattern, the mean differences remaining constant over all 
magnitudes. Thus our catalogue can be considered to have been successfully 
placed on the system ICRS, as defined by Hipparcos, via Tycho-2.
Note that this is not inconsistent with our earlier discovery of trends
in the proper motion differences with Tycho-2.  It simply means
that applying our proper motions to our 1900-epoch positions and bringing
them into a modern epoch will yield modern postions that do disagree with
Tycho-2.  However, it is the integrity of our proper motions that is the
highest priority.  We have little other choice but to adopt the Tycho-2
positional system, at the epoch 1900, for our positional reference system.

\begin{figure}[h!]
 \centering
 \includegraphics[width=0.45\textwidth]{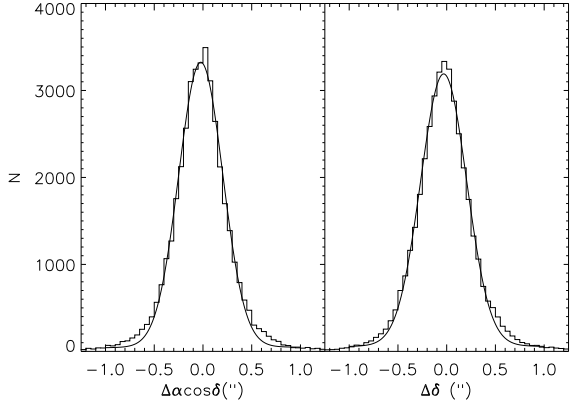}
 \caption{Reference star residuals of CdC-SF catalogue based on a comparison
 with the Tycho-2 positions at the epoch of the plates. We obtain a
 well defined Gaussian distribution, the standard deviations of which are
 $\sigma_{\alpha cos\delta}=0\farcs22$ and
 $\sigma_{\delta}=0\farcs24$.}
 \label{fig08}
 \end{figure}

Figure~\ref{fig08} shows that the dispersion of 
these positional differences with Tycho-2 is
$(\sigma_{\alpha cos \delta},\sigma_{\delta})=(0\farcs22,0\farcs24)$
and is well described by a Gaussian distribution.
This agrees with the mean value obtained for the standard deviation 0$\farcs$34
(final line in the Table~\ref{table01}), that it is
1.414 times the spread of a Gaussian function.

We note that our uncertainty values compare well with results from
other studies based on similar plate material,
but digitized with precision measuring machines specialized for 
photographic plates (Table~\ref{table02}, repeated here from Paper~I).

\begin{table}[h!]
\caption{Comparison of various astrometric studies involving CdC plate material.}
\label{table02}
\centering
\begin{tabular}{llccl}
 \noalign{\smallskip}
 \hline\hline
 \noalign{\smallskip}
Reference         &  Machine  &  Precision     &  Accuracy   & $N_{pl}$ \\
 \noalign{\smallskip}
 \hline
 \noalign{\smallskip}

This paper        &  Scanner  &  $0\farcs18$   & $0\farcs20$ & 400 \\
Rapaport 2006    &  APM      &  $0\farcs15$   & $0\farcs20$ & 512 \\
Ortiz-Gil 1998    &  PDS      &  $0\farcs15$   & $0\farcs15$ & 1 \\
Geffert   1996    &  MAMA     &  $0\farcs15$   & $0\farcs20$ & 2 \\
 \noalign{\smallskip}
 \hline
 \noalign{\smallskip}
\end{tabular}
\end{table}

\subsection{Internal and external errors in proper motions}
\label{err}

In cases where the proper motion is calculated 
using the positions of several catalogues of different epochs,
one can estimate the proper-motion uncertainties from
the dispersion of the individual data fittings.
But in the investigation of the CdC-SF we have compared 
the positions only with one more catalogue, UCAC2. This means
we have the adjustment of a straight line to two points.
Nevertheless, for each star we have included in the catalogue an estimate 
of the uncertainties of the proper motions
obtained from the quadrature sum of the stated positional uncertainties 
in the catalogue CdC-SF and in the UCAC2 divided 
by the epoch difference.

\begin{figure}[h!]
\centering
\includegraphics[width=0.45\textwidth]{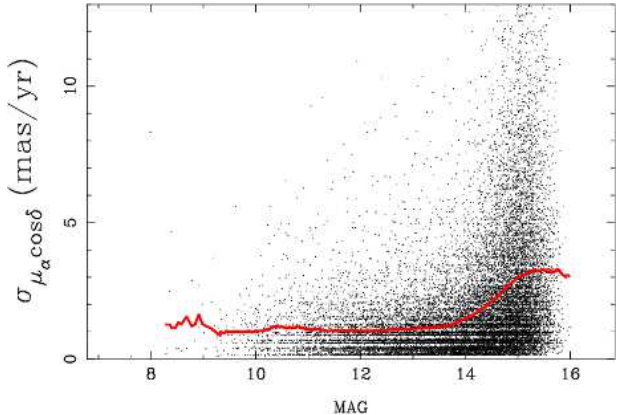}
\caption{Proper-motion uncertainties as a function of magnitude.
Each point represents 25 stars, and the grey line indicates a moving mean.
$\sigma_{\mu_\delta}$ shows a similar behaviour.}
\label{fig09}
\end{figure}

The average of the uncertainties in the two proper-motion
components is (2.0, 1.9)~mas/yr
for all the stars in the catalogue. For stars brighter than V$\leq$14 
it is (1.2, 1.1)~mas/yr, as is shown in Fig.~\ref{fig09}.
The mean uncertainties tabulated by magnitude bins of 
$\Delta mag=1$ are given in Table~\ref{table03}.

\begin{table}[h!]
\centering
\caption{Internal estimation of the proper motion errors
binned by magnitude, including the number of stars per magnitude interval.
Each bin $m$ includes stars whose magnitude is in the range of
$[m-0.5,m+0.5)$.}
\vspace{0.2cm}
\begin{tabular}{rrrr}
 \hline
 \hline
 \noalign{\smallskip}
Mag. & $ <\sigma_{\mu_{\alpha}cos\delta}> $&
$ <\sigma_{\mu_{\delta}}>$  & $N_{*'s}$ \\
      &\multicolumn{1}{c}{(mas/yr)} &\multicolumn{1}{c}{ (mas/yr) }& \\
 \noalign{\smallskip}
 \hline
 \noalign{\smallskip}

   7 &  1.2 & 1.2 &      7\\
   8 &  1.3 & 1.0 &    130\\
   9 &  1.1 & 1.0 &   1092\\
  10 &  1.1 & 1.1 &   5936\\
  11 &  1.1 & 1.1 &  20405\\
  12 &  1.1 & 1.0 &  44464\\
  13 &  1.1 & 1.1 &  77505\\
  14 &  1.6 & 1.4 & 136028\\
  15 &  2.9 & 2.6 & 202416\\
  16 &  3.2 & 3.1 &  15786\\
 \noalign{\smallskip}
 \hline
 \noalign{\smallskip}
 Total  & 2.0 & 1.9 & 503769\\
 \noalign{\smallskip}
 \hline
 \noalign{\smallskip}
\end{tabular}
\label{table03}
\end{table}

\subsection{External comparison with Hipparcos proper motions}

\begin{figure}[h!]
\centering
\includegraphics[width=0.45\textwidth]{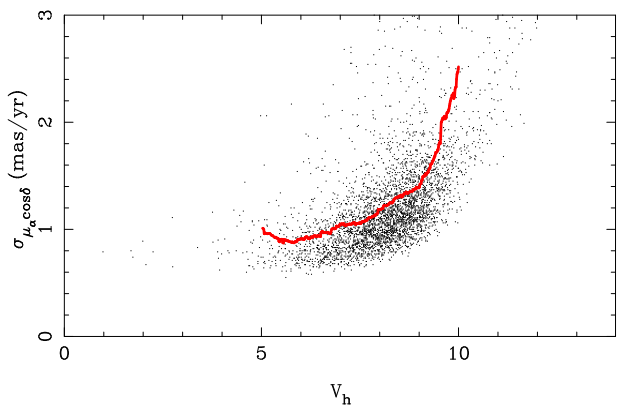}
\caption{Uncertainties of the proper motions as a function of magnitude
from the Hipparcos catalogue (ESA 1997). 
The plot for $\Delta\delta$ is similar.}
\label{fig10}
\end{figure}

The evaluation of the external errors in the proper motions
of the CdC-SF is obtained by means of comparison with the
Hipparcos catalogue (ESA 1997), 
although obviously this comparison can only 
be made for the bright stars.
The uncertainty of the Hipparcos proper motions is on average 
$\sim$1mas/yr (Fig.~\ref{fig10}).

\begin{figure}[h!]
\centering
\includegraphics[width=0.45\textwidth]{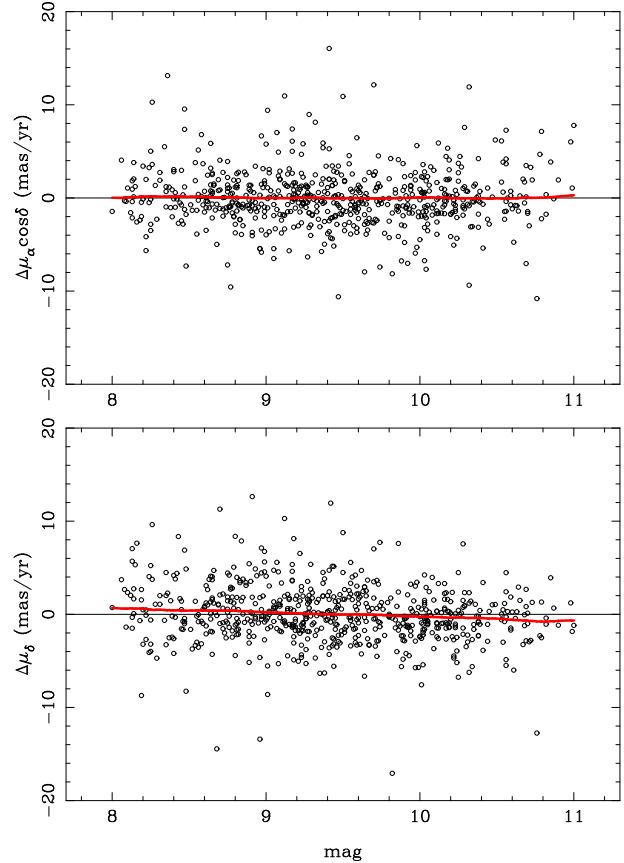}
\caption{Differences of CdC-SF absolute proper motions compared
with Hipparcos as a funcion of magnitude. The fitted line indicates 
the moving-average differences.}
\label{fig11}
\end{figure}

In Fig.~\ref{fig11} 
the differences between our absolute proper motions and 
those of Hipparcos are shown for the 701 common stars.
No systematic trends are seen, neither 
depending on the magnitude nor on the position.

The observed dispersion of the mean differences, 
 $\sigma _ {\Delta\mu _ {\alpha} cos\delta}$~=~2.5~mas/yr and 
 $\sigma _ {\Delta\mu _ {\delta}} $~=~2.7~mas/yr,
includes the errors of our catalogue and 
those of Hipparcos, added in quadrature.
Since the Hipparcos proper motion errors are on average 
1.5~mas/yr, in the common sky area and magnitude range,
we can deduce an estimation of the external accuracy 
of the CdC-SF for the bright objects of up to V$\leq$11, 
of 2.0~mas/yr.

Note that this comparison is done for the brightest stars of our catalogue, 
which are precisely those that suffer from saturation problems, 
but at least it gives us an upper limit of our external errors.

\subsection{Proper motion errors using open clusters}

The internal velocity dispersion of all but the very nearest open clusters 
is too small to be measured with proper motions.
Thus, the observed dispersion is a good estimate
of the individual proper-motion errors for members 
of the cluster.

A handful of open clusters within our catalogue allow us to estimate the
the proper-motion errors and also confirm
that a magnitude equation is not present.

For each cluster utilized, Table~\ref{table04} shows 
the dispersion of our proper motions of the most probable members 
according to the Dias et al (2006) database.
Unfortunately, the membership information available
comes from either bright catalogues 
(Hipparcos, Tycho-2) or from less precise proper-motion sources (UCAC2). 
Thus, some stars may be erroneously identified as members. 
Still, the dispersions of the mean proper motion of the clusters 
provide an estimate of at least an upper limit to our internal error.
We verify that this limit of the error
is consistent with the estimation we made earlier (Sect.~\ref{err}).

\begin{table}[h!]
\centering
\caption{Dispersion of proper motions for each open cluster studied.
Number of stars and magnitude range is also given.}
\vspace{0.2cm}
\begin{tabular}{l|rrrr}
\hline
\hline
\noalign{\smallskip}
Cluster          &  $N_{*s}$ & $\sigma_{\mu_{\alpha}cos\delta}$
&  $\sigma_{\delta}$  & $V_{range}$\\
\hline
\noalign{\smallskip}
 ALESSI 21 &      59  &  2.4 & 2.6  &   9.8  -   12.6\\
 ASCC 24   &      13  &  2.1&  3.4  &   7.5  -   11.7 \\
 ASCC 27   &      13  &  2.6&  2.1  &   8.8  -   12.1\\
 ASCC 30   &      33  &  1.5&  1.8  &   8.9  -   12.2\\
 BOCHUM 3  &       5  &  1.5&  0.7  &  11.9  -   13.7\\
 NGC 2184  &       8  &  2.6&  2.1  &   9.6  -   12.2 \\
 NGC 2215  &       9  &  0.8&  0.9  &  10.6  -   12.9\\
 NGC 2232  &       9  &  1.2&  2.3  &   7.8  -   10.6\\
 NGC 2286  &      31  &  2.4&  2.0  &   9.7  -   14.3\\
 NGC 2302  &      20  &  0.8&  2.1  &  10.8  -   12.5\\
 NGC 2306  &      11  &  1.4&  2.3  &  10.1  -   12.5\\
 NGC 2311  &      14  &  1.4&  1.4  &  10.2  -   12.6 \\
 NGC 2323  &     126  &  1.1&  1.3  &   8.9  -   13.6 \\
 NGC 2335  &       8  &  2.1&  1.1  &  10.7  -   12.7 \\
 NGC 2548  &      86  &  1.1&  1.4  &   8.5  -   15.6\\
\noalign{\smallskip}
\hline
\end{tabular}
\label{table04}
\end{table}

\section{Description of the CdC-SF Catalogue}

     The resulting catalogue, called {\it CdC-SF Catalogue}, 
contains positions and proper motions for approximately 500,000 stars 
in the ICRS system.
It is constructed from 420 plates, one third of the full CdC-SF collection.
The mean epoch is 1901.4, although the positions are presented in the epoch 
1900 for convenience.  The mean epoch of observation for each star is also
listed.  Thus, the original-epoch position of each star can be retrieved
from the proper motions, should one wish to include these data in the 
compilation of other proper-motion catalogues.

\begin{figure}[h!]
\centering
\includegraphics[width=0.45\textwidth]{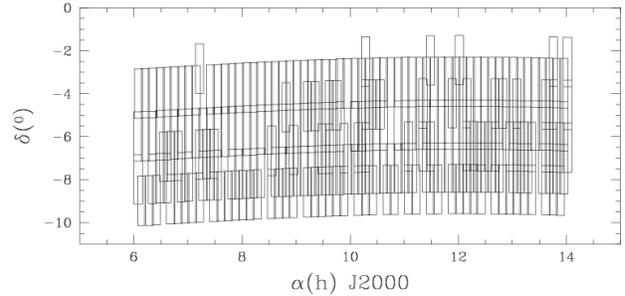}
\caption{Sky coverage of the CdC-SF Catalogue.}
\label{fig12}
\end{figure}

The CdC-SF sky coverage is illustrated in Fig.~\ref{fig12}.
The gaps correspond to plates that had problems 
during the reduction or that were not included because of their
poor quality and would have adversely affected neighbouring fields in the 
catalogue.
The contents of the CdC-SF Catalogue is given in Table~\ref{table05}, and 
a summary of its main characteristics is described in Table~\ref{table06}.

\begin{table}
\begin{minipage}[t]{\columnwidth}
\caption{Contents of the CdC-SF Catalogue.}
\label{table05}
\centering
\begin{tabular}{ll}
 Col.  1: & running number of this catalogue\\
 Col. 2-4: & right ascension ($^h$ $^m$ $^s$) epoch 1900, Eq=J2000\\
 Col. 5-7: & declination ($^{\circ}$ $'$ $''$) epoch 1900, Eq=J2000\\
 Col.  8: & estimate magnitude \\
 Col.  9: & mean epoch of the measures \\
 Col. 10: & proper motion in right ascension \\
          & $\mu_{\alpha}cos\delta$ (mas/yr) \\
 Col. 11: & proper motion in declination\\
          & $\mu_{\delta}$ (mas/yr) \\
 Col. 12: & mean error of  right ascension ($''$) 
 \footnote{Standard deviation of the images from plates overlapped.}\\
 Col. 13: & mean error of declination ($''$) \\
 Col. 14: & mean error of $\mu_{\alpha}cos\delta$ (mas/yr)
 \footnote{Estimation from the positional precisions
of CdC-SF and UCAC2.}\\
 Col. 15: & mean error of $\mu_{\delta}$ (mas/yr) \\
 Col. 16: & mean error of magnitude \\
 Col. 17: & number of contributing overlapped images\\
 Col. 18: & B magnitude in Johnson system 
 \footnote{99.99 value means that the star is not in Tycho-2.}\\
 Col. 19: & V magnitude in Johnson system \\
 Col. 20: & R magnitude \\
 Col. 21: & J magnitude \\
 Col. 22: & H magnitude \\
 Col. 23: & $K_s$ magnitude \\
 Col. 24: & cross number with HIPPARCOS Catalogue 
 \footnote{Blank space means that the star is not in that catalogue.}\\
 Col. 25: & cross number with TYCHO-2 Catalogue\\
 Col. 26: & cross number with UCAC2 Catalogue\\
 Col. 27: & cross number with 2MASS Catalogue\\
\end{tabular}
\end{minipage}
\end{table}

\begin{table}[h!]
\centering
\caption{CdC-SF Catalogue properties.}
\label{table06}
\vspace{0.2cm}
\begin{tabular}{lr}
\hline \hline
\multicolumn{2}{c}{CdC-SF CATALOGUE}\\
\hline 
mean epoch & 1901.4 \\
system & ICRS \\
area covered  &  $\sim$1080 degrees$^2$\\
position range in $\alpha $& $06^h \leq \alpha \leq 14^h$ \\
position range in $\delta$ & $-10.5^{\circ} \leq \delta \leq -2.5^{\circ}$  \\
magnitude range &  $6 \leq  V \leq 16.3$ \\
completeness &  $V \simeq 15.1$ \\
number of stars   &   503769 \\
Hipparcos stars  &  701 \\
Tycho-2 stars   & 40548 \\
\noalign{\smallskip}
measuring error & $3\ \mu m \sim 0\farcs18 $\\
positional error & ($0\farcs21,0\farcs19$)  \\
\qquad \qquad \qquad    (V$<$14)       & ($0\farcs12,0\farcs11$)  \\
$\mu$ error  (mas/yr) & (2.0,1.9)  \\
\qquad \qquad \qquad   (V$<$14)       &   (1.2,1.1)  \\ 
\noalign{\smallskip}
\hline
\end{tabular}
\end{table}

In the final catalogue we have rejected stars that come from just one measure,
because, at minimum, every real stellar image should have at least two
measures owing to the fact that we have digitized every plate twice. 
The two scans of the same plate were considered as separate measures, 
as a justifiable simplification. 
Even if those two scans are not truly independent, 
they are still significantly enough independent because 
of the dominant contribution of the measuring process to the total error.
So, we think that treating them as independent measures 
for the purposes of computing averages and uncertainties might be roughly correct.  
As far as the uncertainty estimates goes, the validity of this approximation is shown 
to be reasonable by the external checks on the error estimates, i.e, 
the open clusters and the comparison with the Hipparcos data.
We have also removed uncertain or failed matches with stars in the UCAC2,
which were needed to calculate proper motions. For this reason, the UCAC2 completeness
sets our completeness. 
In addition to this we have further losses due
to the detection pipeline in combination with the problematical plates.

Besides the photometric magnitude calculated and calibrated during 
the reduction, we have included the photometry from other catalogues:
$BV$ magnitudes for the common stars with Tycho-2, 
$R$ magnitude from UCAC2 and $JHK$ from the 2MASS catalogue.
Tycho-2 magnitudes, $B_T$ and $V_T$, have been transformed into the
the UBV photometric system using the equations (ESA SP-1200 1997):
 
\begin{center} 
\begin{equation}
\begin{array}{lcl}
V &=& V_T -0.090\ (B_T-V_T)\\
B-V &=& 0.850\ (B_T-V_T)
\end{array}
\end{equation}
\end{center}

\begin{figure}
\centering
\includegraphics[width=0.4\textwidth]{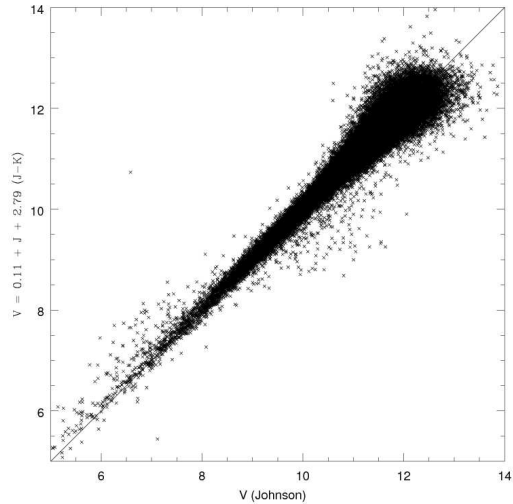}
\caption{Correlation between the $V_{pseudo}$ estimate and Hipparcos magnitude.
The straight line $y=x$ confirms that
the approximation is adequate enough for our purposes.}
\label{fig13}
\end{figure}

\begin{figure}[h!]
\centering
\includegraphics[width=0.45\textwidth]{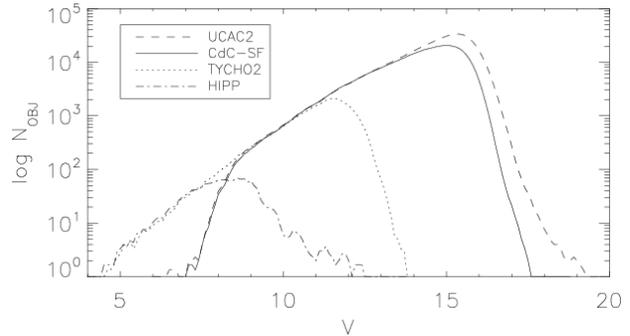}
\caption{Completeness of the CdC-SF catalogue:
magnitude distribution compared to other astrometric catalogues.}
\label{fig14}
\end{figure}

\begin{figure}
\centering
\includegraphics[width=0.45\textwidth]{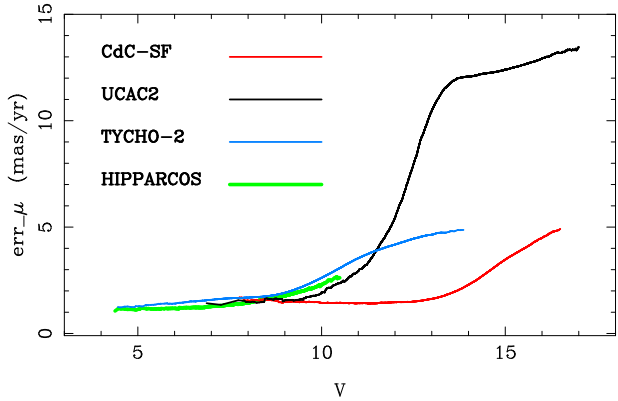}
\caption{Moving-average error of CdC-SF proper motions,
as a function of magnitude, compared with Hipparcos, Tycho-2 and UCAC2.}
\label{fig15}
\end{figure}

An extrapolated $V$ estimate has been derived 
from 2MASS $J,K$ photometry, so-called pseudo-$V$, following
the empirical relation given by Girard et al.~(2004).
This approximation works reasonably well over a range of spectral types
(see Fig.~\ref{fig13}) and allows us to compare the magnitude distributions 
of several catalogues on a common system.
Figure~\ref{fig14} shows the magnitude
distribution of the CdC-SF catalogue with other astrometric catalogues
for the same sky coverage, repeated here from Paper~I for convenience.
The falloff of the CdC-SF distribution is at V=15.1.
Figure~\ref{fig15} shows the distribution of proper-motion errors
of the CdC-SF catalogue as a function of magnitude
in comparison with other proper-motion catalogues.

The catalogue includes also cross-identifications with  
the Hipparcos, Tycho-2, UCAC2 and 2MASS catalogues in order 
to access complementary information from  
these catalogues. 
We are aware that CdC-SF does not include all Hipparcos stars in our area
(bright objects are saturated on the plates), nor
all Tycho-2 stars (lost on the grid or on rejected plates).
For this reason, we annex two files with the identification numbers 
of these missing stars, 3170 Hipparcos's stars and 39611 
stars of Tycho-2 not included in Hipparcos.

Future versions of the CdC-SF catalogue with expanded coverage to the full 24h
of right ascension will be constructed by applying the techniques presented here
to the remainder of the digitized plate material.

\section{Conclusions}

We have presented the CdC-SF Catalogue, an astrometric catalogue
with positions and proper motions for more than 500,000 stars, 
in the ICRS system, with a mean epoch of 1901.4,
to $V=16$, covering 1080 degrees$^2$ in the sky area
between $\alpha=(06^h,14^h),\delta=(-10.5^{\circ},-2.5^{\circ})$.
The mean uncertainty in positions is 0$\farcs$20 
(0$\farcs$12 for well-measured stars), 
and in proper motions it is 2~mas/yr (1.1~mas/yr for well-measured stars).
The catalogue is based on the reduction of the original
{\it Carte du Ciel} photographic plates, San Fernando zone,
that have been ``revived'' by digitization with a commercial flatbed scanner.

External comparison of the data has been
made with the catalogues Hipparcos and UCAC2,
as well as a study of the proper-motion errors using open clusters. 
The catalogue achieves approximately the same precision in proper motions
as the Hipparcos catalogue, but it extends to seven magnitudes fainter.

Finally, the CdC-SF catalogue covers -10$^{\circ}$ to +60$^{\circ}$ 
galactic latitude and will provide an excellent basis for studies 
of the kinematics of different components of our Galaxy.
The positions at the epoch 1900 should prove to be useful for 
other proper-motion determinations.

\begin{acknowledgements}
We are very grateful to the Observatorio de San Fernando for
making available to us the Carte du Ciel plates from their historical archive.
We also want to thank all of the people who have participated in
the digitization of the collection, with special
mention of Jos\'e Mui\~nos, Fernando Beliz\'on and Miguel Vallejo.
\end{acknowledgements}

\end{document}